\documentclass[aps,pra,twocolumn,groupedaddress,floatfix,eqsecnum,nofootinbib,showpacs]{revtex4}
\usepackage{amsmath}
\usepackage{amssymb}
\usepackage{graphicx} 
\usepackage{psfrag} 
\usepackage{bm} 
\input colordvi

\begin{document}

\newcommand{\Sref}[1]{Section~\ref{#1}}
\newcommand{\sref}[1]{Sec.~\ref{#1}}
\newcommand{\Cref}[1]{Chap.~\ref{#1}}
\newcommand{\tql}{\textquotedblleft} 
\newcommand{\tqr}{\textquotedblright~} 
\newcommand{\tqrc}{\textquotedblright} 
\newcommand{\Refe}[1]{Equation~(\ref{#1})}
\newcommand{\Refes}[1]{Equations~(\ref{#1})}
\newcommand{\fref}[1]{Fig.~\ref{#1}}
\newcommand{\frefs}[1]{Figs.~\ref{#1}}
\newcommand{\Fref}[1]{Figure~\ref{#1}}
\newcommand{\Frefs}[1]{Figures~\ref{#1}}
\newcommand{\Tref}[1]{Table~\ref{#1}}
\newcommand{\reff}[1]{(\ref{#1})}
\newcommand{\refe}[1]{Eq.~(\ref{#1})}
\newcommand{\refes}[1]{Eqs.~(\ref{#1})}
\newcommand{\refi}[1]{Ineq.~(\ref{#1})}
\newcommand{\refis}[1]{Ineqs.~(\ref{#1})}
\newcommand{\framem}[1]{\overline{\overline{\underline{\underline{#1}}}}}
\newcommand{\PRA }{{ Phys. Rev.} A }
\newcommand{\PRB }{{ Phys. Rev.} B} 
\newcommand{\PRE }{{ Phys. Rev.} E}
\newcommand{\PR}{{ Phys. Rev.}} 
\newcommand{\APL }{{ Appl. Phys. Lett.} }
\newcommand{\PRL}{Phys.\ Rev.\ Lett. }
\newcommand{\OCOM }{{ Opt. Commun.} } 
\newcommand{\JOSA }{{ J. Opt. Soc. Am.} A}
\newcommand{\JOSB }{{ J. Opt. Soc. Am.} A}
\newcommand{\JMO }{{J. Mod. Opt.}}
\newcommand{\RMP}{Rev. \ Mod. \ Phys. }
\newcommand{\etal} {{\em et al.}}

\title{High-Q Optical Cavities in Hyperuniform Disordered Materials}
\author{Timothy Amoah}
\author{Marian Florescu} \email[Electronic Address: ]{ m.florescu@surrey.ac.uk}

\affiliation{Advanced Technology Institute and Department of Physics, University of
  Surrey, Guildford, GU2 7XH, United Kingdom}

\date{\today}

\begin{abstract}
We introduce the first designs for high-Q photonic cavities in slab  architectures in hyperuniform disordered solids displaying  isotropic band gaps. Despite their disordered character, hyperuniform disordered structures have the ability to tightly confine the TE-polarised radiation in slab configurations that are readily fabricable.  The architectures are based on carefully designed local modifications of otherwise unperturbed hyperuniform dielectric structures. We identify a wide range of confined cavity modes,  which can be classified according to their approximate symmetry (monopole, dipole, quadrupole, etc.) of the confined electromagnetic wave pattern. We demonstrate that quality factors ($Q$) $Q>10^{9}$ can be achieved for purely 2D structures, and that for three--dimensional finite-height photonic slabs, quality factors  $Q>20,000$ can be maintained. 
                   
\end{abstract}

\maketitle

A special class of disordered photonic heterostructures has recently been shown to display large isotropic band gaps comparable in width to band gaps found in photonic crystals \cite{pnas_flo, pnas_flo_2,opt_express}. The large band gaps found in these structures are facilitated by the hyperuniform geometrical properties of the underlying point-pattern template upon which the structures are built. The statistical isotropy of the photonic properties of these materials is highly relevant for a series of novel photonic functionalities including arbitrary angle emission/absorption and free-form wave-guiding \cite{opt_express}, \cite{vynck_natmat}. 

A point pattern in real space is hyperuniform if for large $R$ the number variance $\sigma(R)^2$  
within a spherical sampling window of radius $R$ (in $d$ dimensions), 
grows more slowly than the window volume, i.e., more slowly than $R^d$.  In Fourier space, hyperuniformity means that the structure factor $S({\bf k})$ approaches zero as $|{\bf k}| \rightarrow 0$. \cite{Torquato2003,sal_Acoeff}. The lack of periodicity in these hyperuniform disordered solids (HUDS) demonstrates that Bragg scattering is not a prerequisite for photonic band gaps and that interactions between local resonances and multiple scattering are sufficient,  provided that the disorder is constrained to be hyperuniform \cite{pnas_flo}.

The concept of optical cavities in hyperuniform disordered (HUD) photonic materials was recently introduced in Ref.~\cite{tm-wg_flo}. The structure analysed was obtained by placing dielectric rods at each point of a hyperuniform point pattern. The radius of one selected rod was varied to achieve localisation of the TM-polarised electromagnetic field at that point. The study was based purely on 2D structures and vertical confinement, the primary loss pathway in real slab structures, was not discussed. The question of how to achieve index-guiding and the essential vertical confinement in disordered photonic slab structures, a prerequisite of realising cavities with high-quality factors in fabricable structures, was seen as a potential roadblock in the HUD photonic materials field.

In this paper we discuss analogous network structures for localising TE-polarised radiation and analyse the vertical confinement issue. The structures analysed are composed of polygonal cells of dielectric walls. The protocol for generating these structures described in Refs.~\cite{pnas_flo},\cite{prb_qc} consists of Delaunay triangulating over a hyperuniform point pattern  and connecting the center of mass of the Delaunay triangles to form cells that contain one original point each. The walls of the resulting network are then given a finite thickness. Trivalency is preserved throughout the structure. The structure presents well defined short-range order and long-range statistical isotropy.
The resulting structure can be described as a hyperuniform disordered honeycomb (HDH). 
Recently, these designs have been fabricated on the microwave scale and successfully tested
\cite{Man2014,Man2014a}.   

\begin{figure}[h]
  {\centerline{\includegraphics*[width=0.8\linewidth]{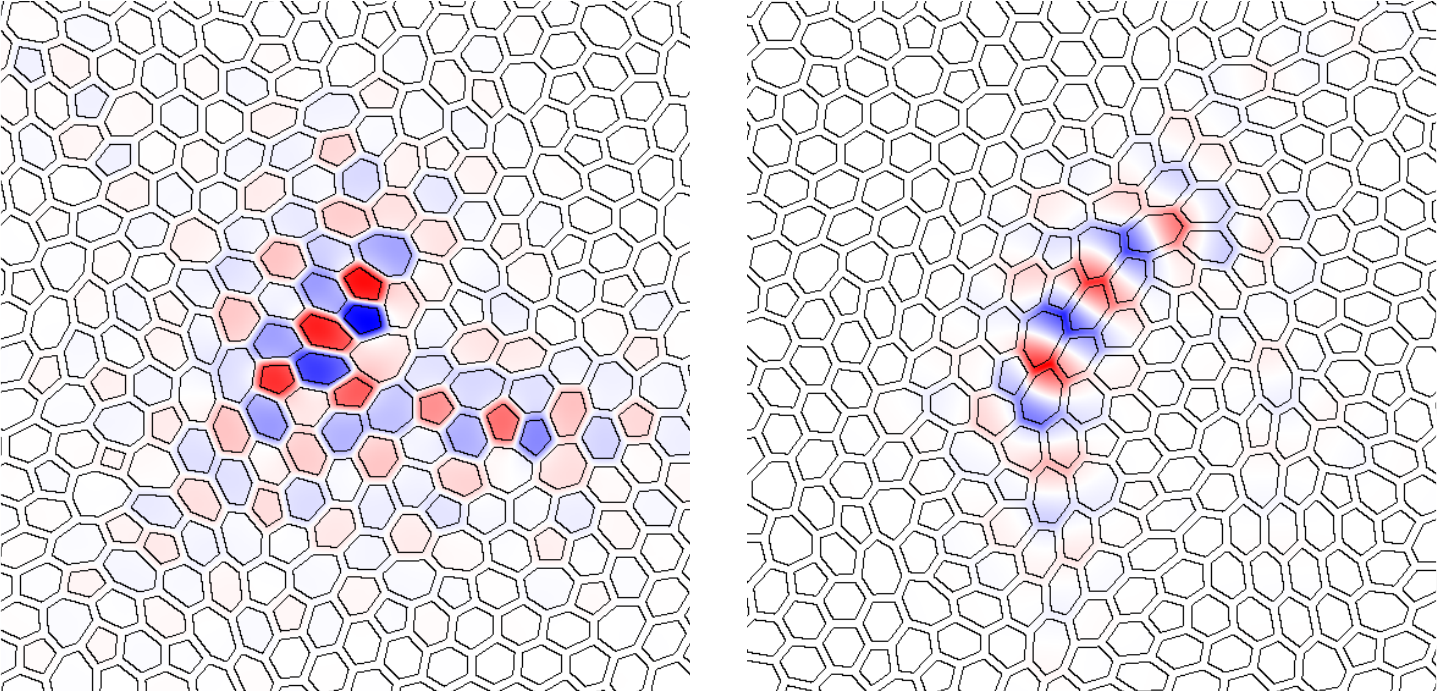}}}
  \caption{ (Color online) Magnetic field distributions of two Anderson-like localised modes, (left) below the lower and (right) above the upper photonic band gap edge respectively.} 
  \label{fig:anderson}
\end{figure}

%\section{2D Simulation}

We begin our analysis with the study of purely 2D structures. A length scale $a = L/\sqrt{N}$ is defined, such that an $N$-point hyperuniform pattern in a square box of side length $L$ has density of $1/a^2$. The structure parameters are set to $\epsilon=11.56$ for the dielectric constant and the wall width is set to $w=0.2a$. For our study, we employ a point pattern containing $N=500$, with a hyperuniformity order parameter $\chi=0.5$ \cite{hyper_stealthy}. We calculate the photonic band structure using the eigenmode expansion software ``MPB" for a $\sqrt{N} a \times \sqrt{N} a$ sample and identify the photonic band edges. We then run a sweep for varying wall thickness ($w$) and find the largest band gap with $\Delta \omega / \omega_C = 30.7\%$ for $w=0.23a$, where $\Delta \omega$ is the band gap width and $\omega_C$ is the band gap center frequency.  Similar to other conventional disordered structures, unperturbed HUD structures display defect modes. These Anderson-like localised modes,  shown in \Fref{fig:anderson}, occur naturally at the photonic band gap edges and extend over over network domains comprising 5-10 cells. Sparsely occurring are accidental localised modes extending on 1-3 network cells which are promoted into the photonic band gap topologically, but we are not exploring these here \cite{apl_dis}.  

\begin{figure}[ht]
{\centerline{ \includegraphics*[width=0.8\linewidth]{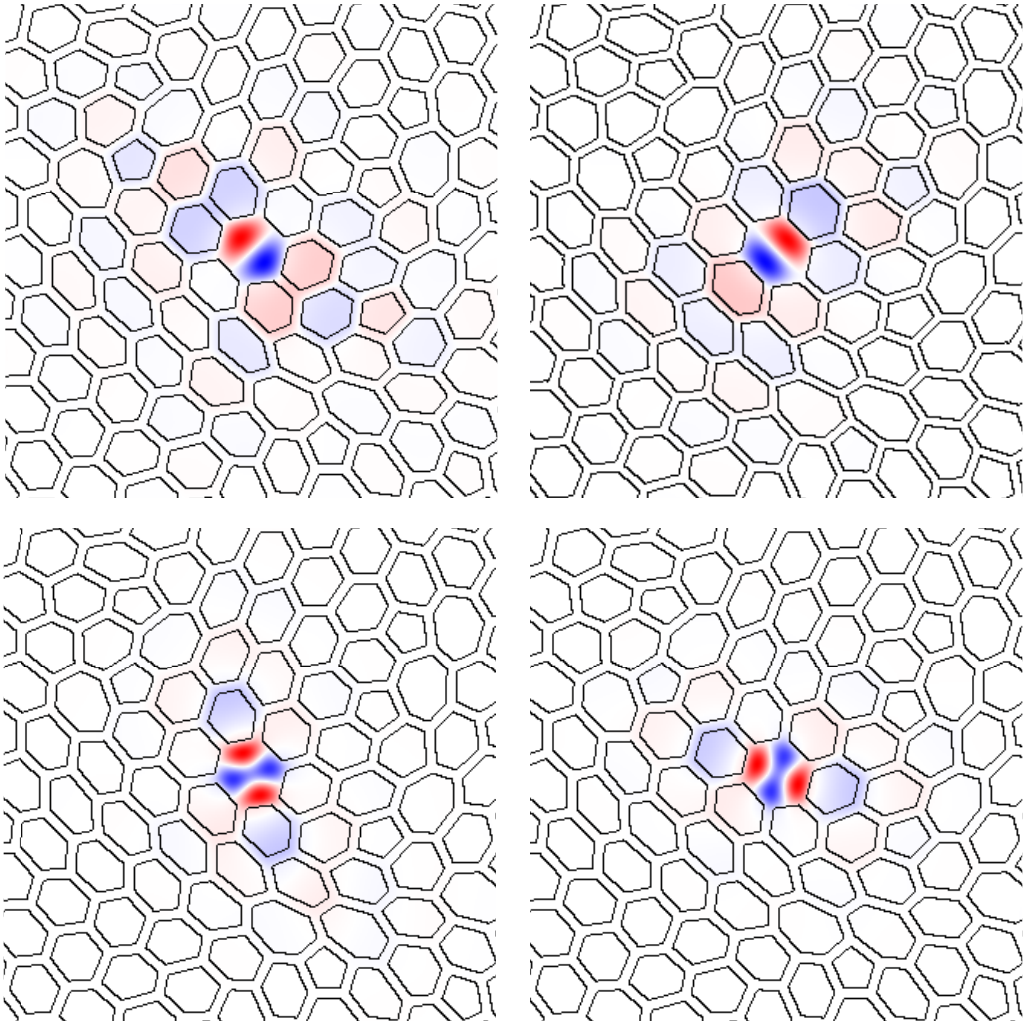}}}
\caption{ (Color online) Magnetic field distribution for the four defect modes in the unmodified H1 cavity in a structure of wall thickness $w=0.2a$. The modes are labelled  D$_1$, D$_2$, Q$_1$, Q$_2$ (shown from left to right in the figure).} 
\label{fig:cavity_thin_walls}
\end{figure}

In an otherwise unperturbed HDH structure, it is possible to create an intentional localized state of the electromagnetic field by reducing or enhancing the dielectric constant at a certain point in the structure. 
For a triangular lattice of holes it is common practice to fill a single hole to make a cavity which is often labelled a H1 cavity. In analogy to this we fill a single cell and also label it H1. Due to the presence of the defect, four localized cavity modes are created within the photonic band gap at specific frequencies. The mode profiles are shown in \Fref{fig:cavity_thin_walls}. Two of the modes are dipole-like and the other two modes are quadrupole-like.
We denote the lower frequency dipole mode D$_1$ and the higher frequency dipole mode D$_2$.
The cavity has an approximately hexagonal shape, as such we can describe the modes according to the approximate symmetries associated with a hexagonal unit cell in a triangular lattice.
For the D$_1$ mode the nodal axis of the $H_{z}$ component of the electromagnetic field lies approximately along the faux K direction and the mode is propagation confined along the faux M direction. For the D$_2$ mode the nodal axis of the $H_{z}$ component of the electromagnetic field lies approximately along the faux M direction and the mode is propagation confined along the faux K direction. The magnetic field distribution of the cavity modes shown in  \Fref{fig:cavity_thin_walls} suggests very good in-plane confinement and calculations of the quality factors (using the finite difference time domain software ``MEEP" \citep{meep}) confirms that the quality factor for all modes higher than $10^{9}$.

\begin{figure}[ht]
{\centerline{ \includegraphics*[width=1.00\linewidth]{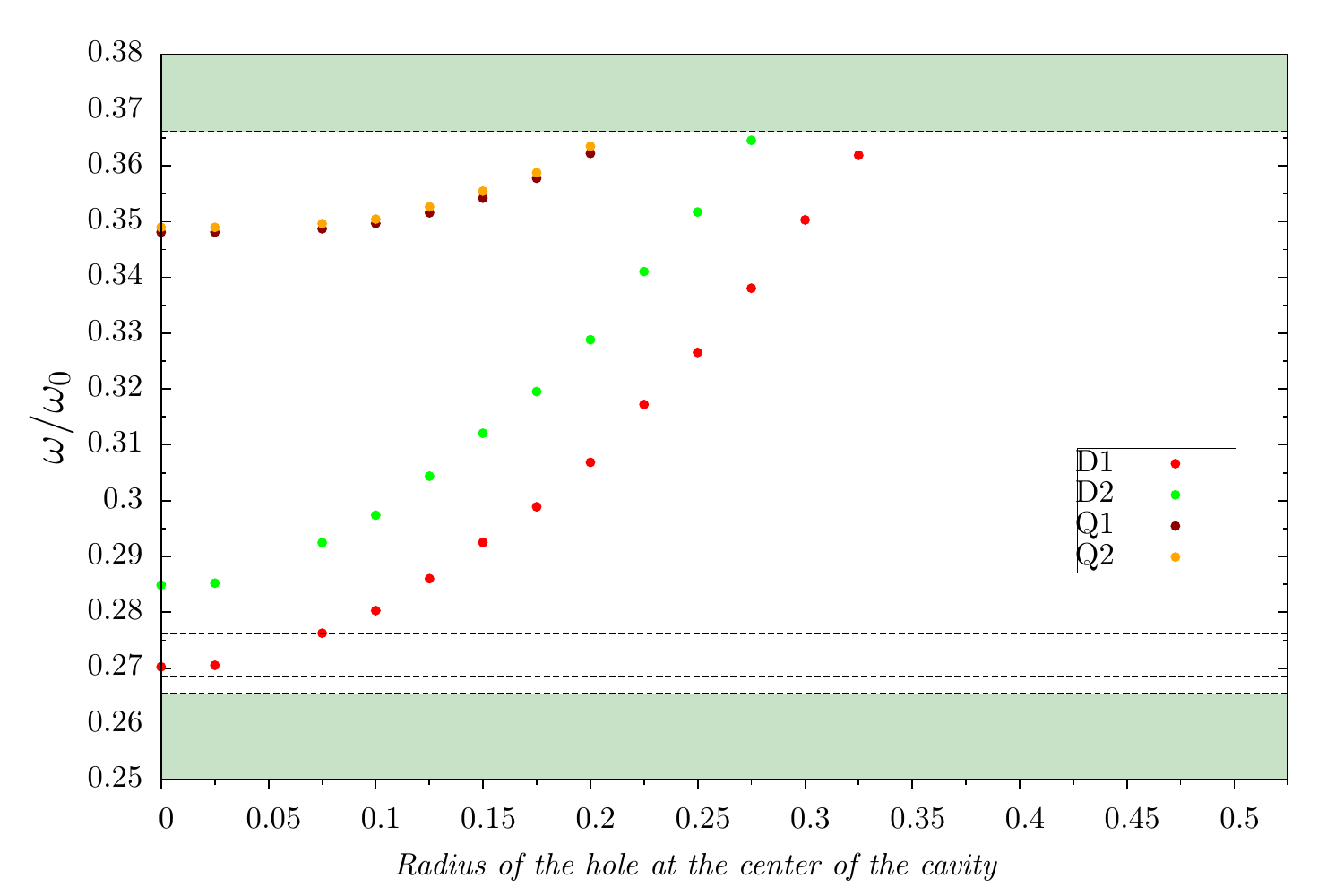}}}
\caption{ (Color online) Frequencies of the cavity modes in the band gap obtained for different radii of a hole placed at the center of the cavity.} \label{fig:radius_thin_walls}
\end{figure}

\begin{figure*} [ht]
\center
\includegraphics*[width=1\textwidth]{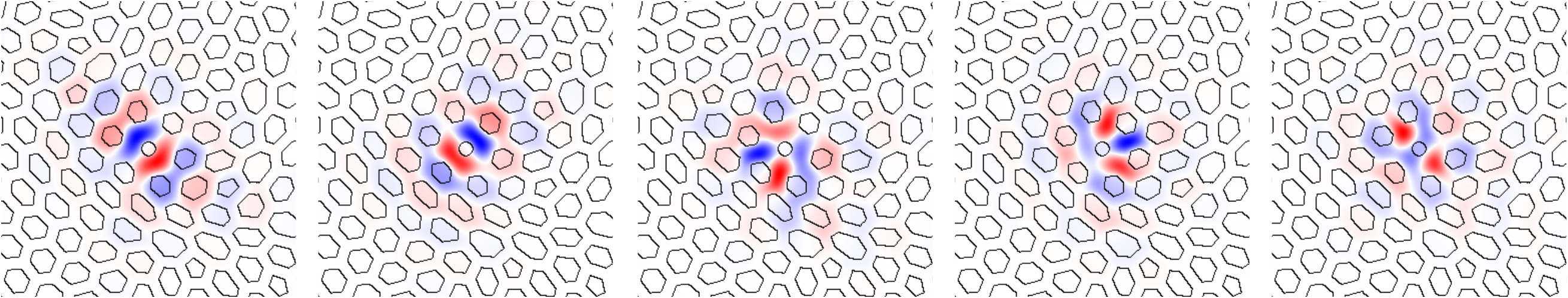}
\caption{ (Color online) Magnetic field distribution for the five cavity modes in the enlarged H1 cavity with center hole $r=0.2a$ in a structure of wall thickness $w=0.4a$ .     
  The modes are labelled  D$_1$, D$_2$, Q, H, O left to right.} \label{fig_cav_fields_1}
\end{figure*}

\begin{figure}[ht]
{\centerline{ \includegraphics*[width=1.00\linewidth]{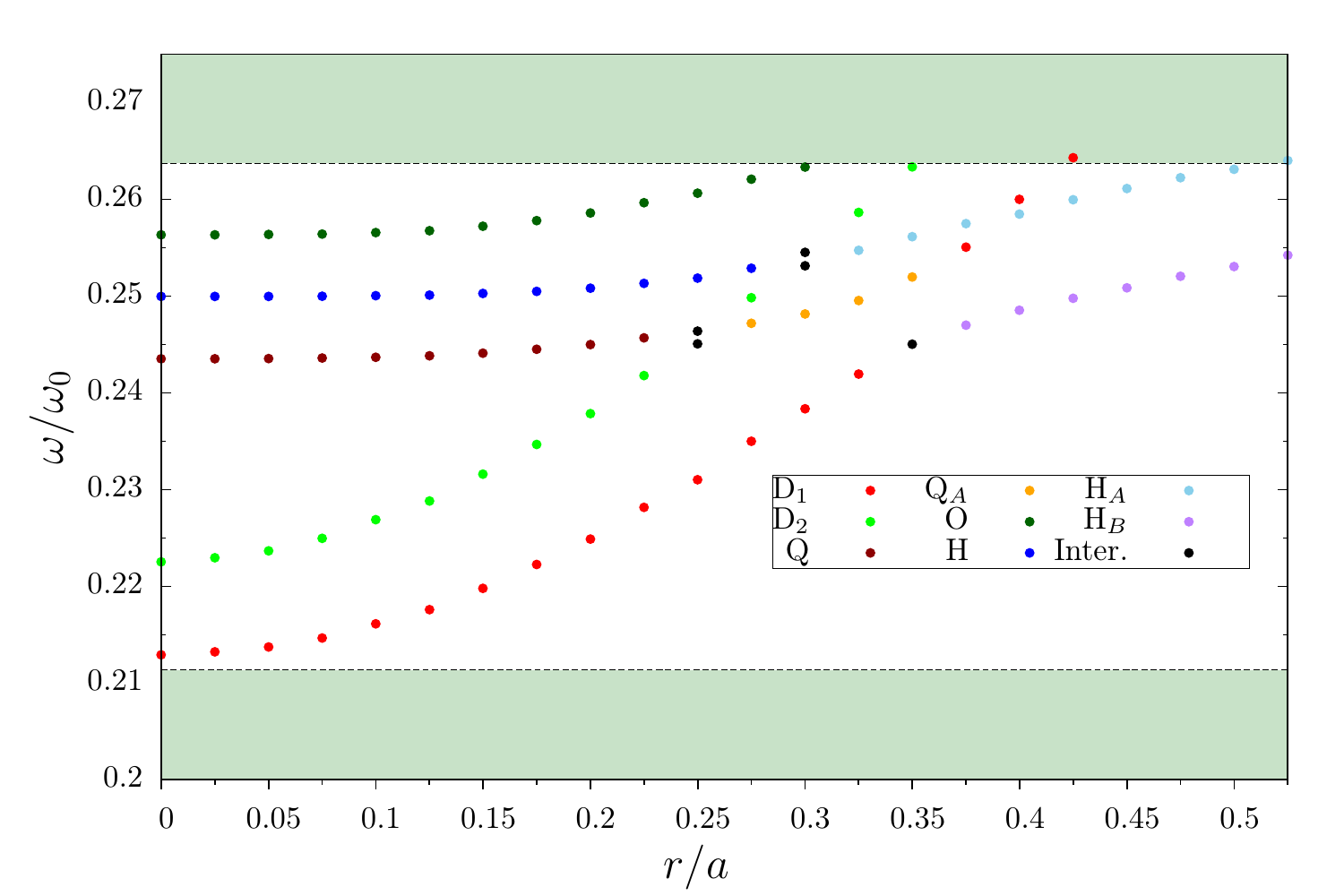}}}
\caption{ (Color online) Frequencies of the cavity modes in the band gap obtained for different radii of a hole placed at the center of the cavity. Modes are labelled according to their number of pole lobes. Different colours are used for modes with the same number of lobes but which have distinct field patterns. Cross over intermediate modes are coloured black.} \label{fig_cav_fields_2}
\end{figure}

\begin{figure}[t]
  {\centerline{\includegraphics*[width=0.95\linewidth]{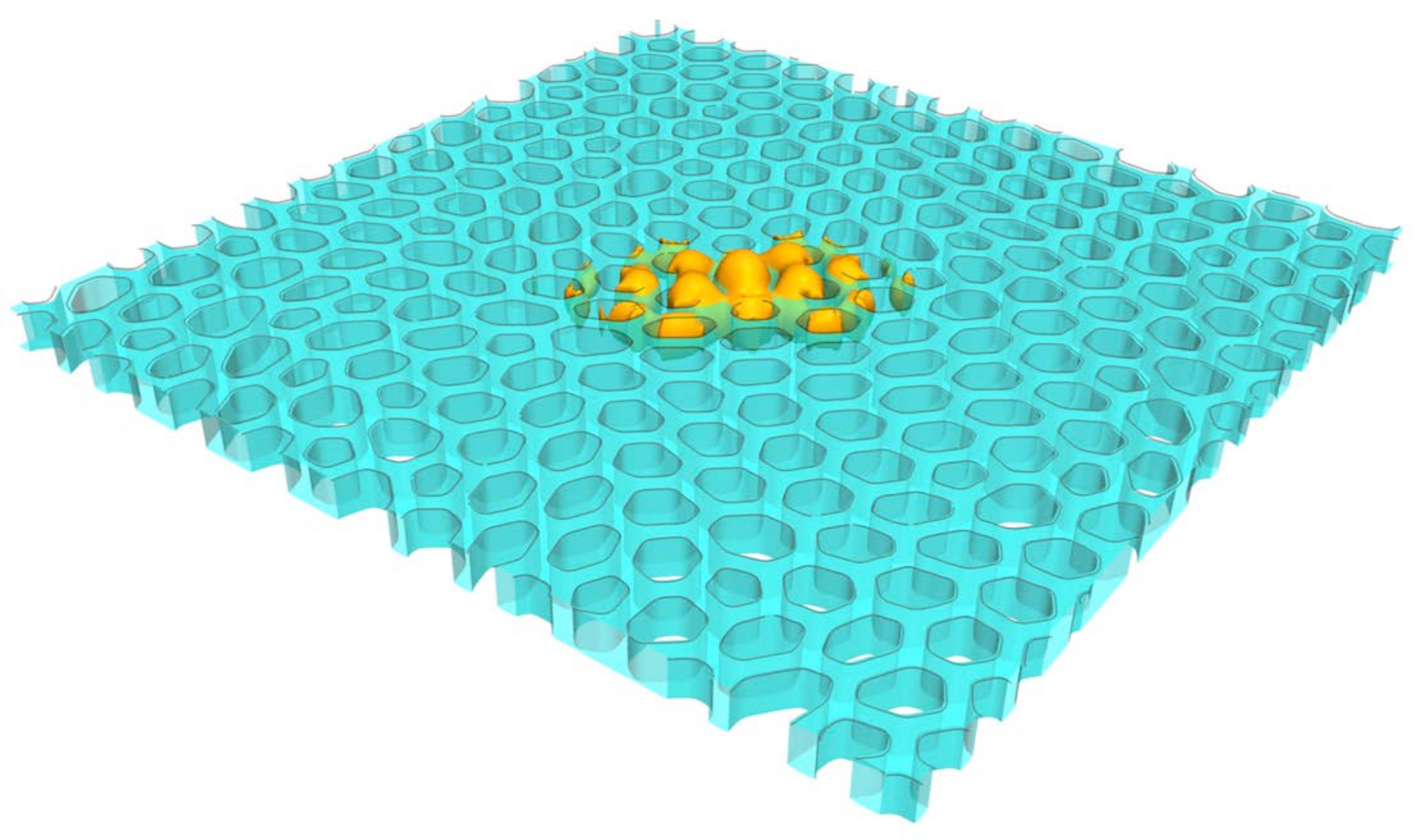}}}
  \caption{ (Color online) Intensity distribution (yellow) of a confined cavity mode in a
  hyperuniform disordered honeycomb (HDH) photonic slab (blue).}
  \label{fig:3Dstructure}
\end{figure}

\begin{table}[h]
\centering
    \begin{tabular}{c|c|c|c}
        %\hline
Mode     & $\omega / \omega_0$ in-plane  & $\omega / \omega_0$ slab & Q-factor        \\ \hline
D$_1$       & 0.22482                     & 0.26722                    & 7976            \\
D$_2$       & 0.23775                     & 0.28003                    & 8672            \\
Q        & 0.24546                     & 0.28811                    & 2561            \\
H        & 0.25133                     & 0.29385                    & 4845            \\
O        & 0.25916                     & 0.29973                    & 3230            \\
    \end{tabular}
\caption{The mode frequencies in the 2D and 3D case and Q factor for the modified H1 cavity with a $0.2 a$ inner hole}
\label{tab:quality}
\end{table}

To study the frequency behaviour of the cavity modes as a function of dielectric filling of the defect cell, we are placing a circular air hole at the center of the unmodified H1 cavity, vary its radius from $r=0$ to $r=0.535a$ and monitor the evolution of the cavity mode frequencies. The general trend is that the frequency of the cavity modes increases with the radius of the hole. In other words, the mode frequency is lowered by the inclusion of more dielectric, which demonstrates that the cavity originates from modes above the upper band edge of the uperturbed structure. The same behaviour is observed in the TM case when the radius of a defect rod is increased \cite{tm-wg_flo}. 

We now turn our investigation towards the photonic slab where we no longer can neglect the vertical extent of the structure. This is the same as the concept of photonic-crystal slabs \cite{pc_slabs,pc_book}, which has proven to be a promising platform for integrated photonic micro-circuitry. In photonic slabs, while there exists no true band gap, the existence of  ``pseudo-band-gaps'' enables low-loss waveguiding  and high Q cavities ~\cite{Johnson1999_gm}.  The ``pseudo-band-gap'' is associated with an effective band gap in the in-plane projected band structure for which modes below the light cone cannot couple to the continuum of states outside the slab. This can also be thought of in terms of ``index confinement" due to the contrast in effective dielectric inside the slab and in the surrounding air region. The projected band structure method only works well when only a single in-plane unit cell is considered.  For supercell calculations, the spectral regions with no effective band gap will be translated onto the region of the effective band gap, through the ``supercell folding'' effect. With no straightforward method to circumvent this we will resort to making the assumptions that a pseudo-band-gap does exist for the photonic slab and that if a cavity mode of specific form lies within the band gap when calculated in 2D it will also lie within the pseudo-band-gap for the photonic slab. Consequently, the frequencies in \Fref{fig:radius_thin_walls} are the in-plane frequencies, not the slab frequencies.

For photonic slabs it is not generally true, that the best optical confinement is achieved 
for the largest in-plane photonic band gap. For a large enough sample the main losses occur in the vertical direction and to prevent them one needs to rely on index confinement. As such, the thicker walls translate into a higher effective index for the slab, and hence, better vertical confinement. For our structures, we use a structure height of $h=0.7a$, and doubled the wall width to $w=0.4a$ to improve confinement. For an unmodified H1 cavity the expected quality factor is typically rather moderate even for a conventional photonic crystal, $Q<500$ \cite{Tandaechanurat2008_H1_thick}. We calculate $Q=210$ for the D$_1$ and $Q=190$ for the D$_2$ mode in the HDH. This values can be considerably increased by placing an inner hole at the centrer of the cavity (see Table~\ref{tab:quality}).

A popular method of of achieving optimal cavity designs is to reduce adjacent holes slightly in size and shift them outwards along the lattice directions. In the HDH case there are no lattice directions, and we instead shift the shrunk cells along the vector given by the center of mass of the cavity to the center of mass of the neighbouring cell. The cells are shrunk to 52\% of the size of a cell with infinitesimal walls and are shifted by 8\% of the length of the vector outwards. For a modified cell cavity we find 5 cavity modes laying in the band gap. The lowest frequency mode is again the D$_1$ mode followed by the D$_2$ mode. An isocontour plot of the electric field intensity distribution of the D$_1$ mode is shown in \Fref{fig:3Dstructure} displaying in-plane PBG-mediated confinement and vertical refractive index confinement. As expected, the two dipole modes have now lower frequencies due to the addition of dielectric material. The next higher frequency modes are a quadrupole-like mode (Q) and a hexapole-like mode (H). Comparing the two mode patterns, we note that (due to disorder) the symmetry separation of the two modes is not entirely complete. Lastly, there is the highest frequency mode which is rather difficult to define. The $H_{z}$ field is symmetric with respect to the faux K direction; therefore the mode is an odd mode. We can identify 4 nodes lying ``within" the cavity, so it can either be considered a second quadrupole mode or an octapole mode. We assert that since it is higher in frequency than the hexapole mode we label it as an octapole-like mode (O).  

\begin{figure}[ht]
{\centerline{ \includegraphics*[width=1.00\linewidth]{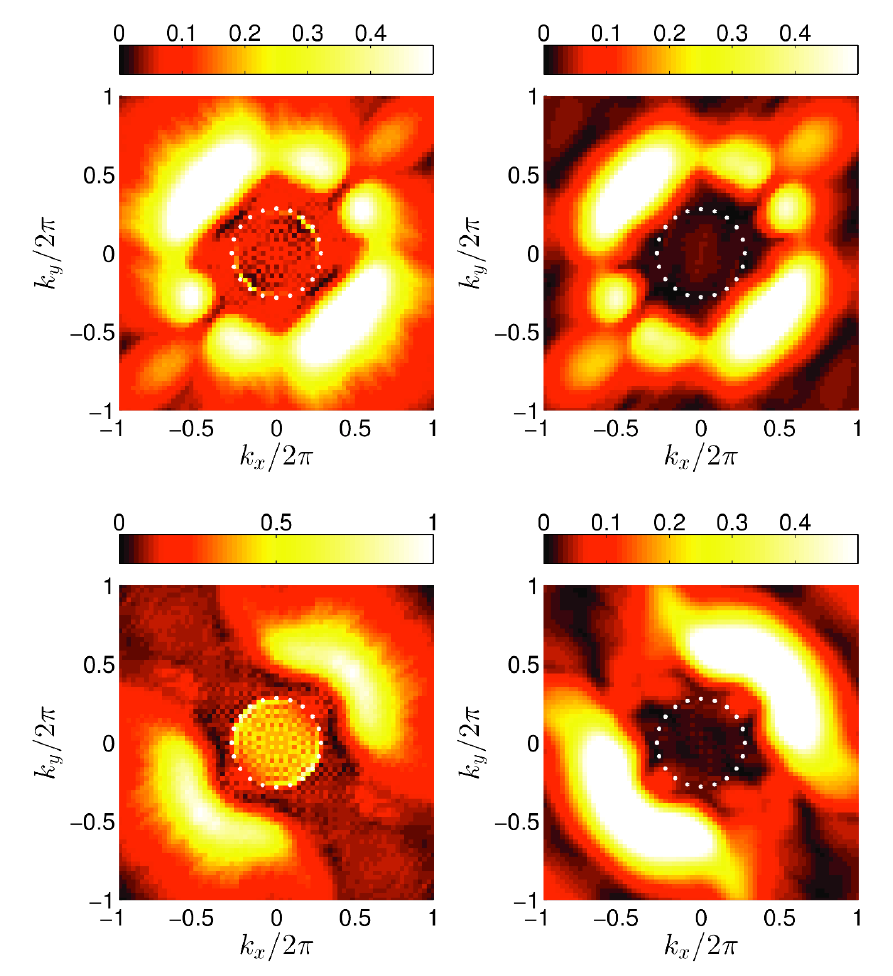}}}
\caption{ (Color online) Spectrum of the Fourier components of the electric field distribution for the D$_1$ mode before (top left) and after modification (top right) and D$_2$ mode before (bottom left) and after modification (bottom right) . Logarithmic colour scale.} \label{fig:fourier}
\end{figure}

\Fref{fig:fourier} shows the spatially Fourier transformed in-plane fields of the D$_1$ and D$_2$ cavity modes for both the unmodified and modified (extra central hole) cavity. The respective light circles are indicated by the white dots. Only the Fourier components with $k<\omega c$ can radiate into the far-field and hence the more Fourier components that are within the light circle the greater the radiative loses in the vertical direction. Clearly, the modified-cavity design in \Fref{fig:fourier} displays a cosniderably smaller number of radiative components.
 
\begin{figure}[ht]
{\centerline{ \includegraphics*[width=1.00\linewidth]{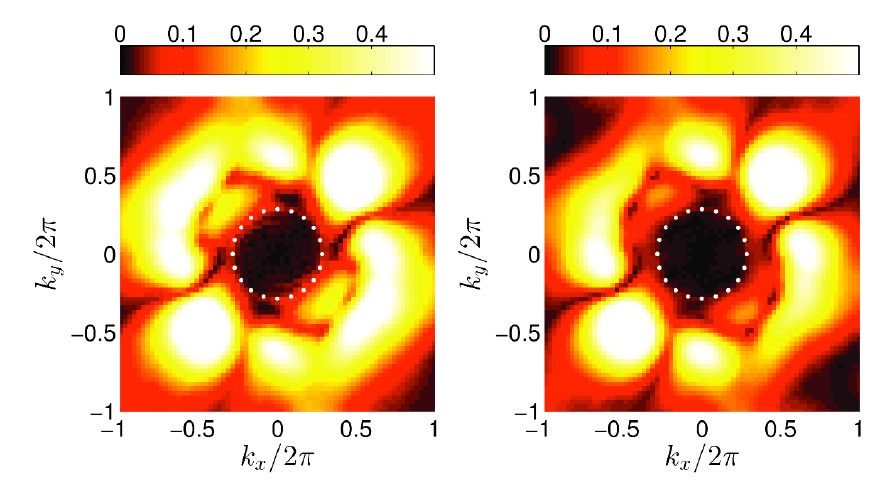}}}
\caption{ (Color online) Spectrum of the Fourier components of the electric field distribution for a modified cavity with neighbour cell size of 52\% (left) and 44\% (right), both shifted by 8\%. Logarithmic colour scale.} \label{fig:ft_O}
\end{figure}

Finally, we consider optimal cavity designs for the O-mode. For a neighbour cell size of 44\% at the same center of mass shift of 8\%,  we find a very high quality factor of $Q=20,148$, which is significantly larger than $Q=3230$ obtained at 52\% neighbour cell size. \Fref{fig:ft_O} shows the Fourier spectrum in each respective case. As before, the Fourier components in the light circle are reduced for the higher quality modification, however the difference in the reduction is less even if a  large increase in quality factor is achieved. Most significant is the change in frequency from 0.29973 to 0.28702, which displace the mode to spectral domain where the radiative losses are minimised.
 
%\section{Summary}

In summary, we have introduced novel architectures for the design of optical cavities in a hyperuniform disordered material.  We have demonstrated that H1-type cavity defects can support
localized modes with a variety of symmetries and multiple frequencies. The ability to localize modes of different symmetry and frequency in the same physical cavity and to guide light through modes with different localization properties can have a great impact on all-optical switching \cite{eu_switch}, implementations of linear-optical quantum information processors \cite{eu_zeno}, and single-photon sources \cite{eu_single_epl,eu_single_jmo}.

In principle, one may have the presumption that disorder would facilitate significant out of plane scattering as compared to periodic structures. Here we have demonstrated that for cavity built on hyperuniform platforms, by adequately adjusting the structure parameters and cavity design it is possible to achieve very tight optical confinement.  

Our successfully demonstration of high-Q cavity for transverse electric (TE) polarized radiation is encouraging future investigation of TE wave-guides in disordered photonic slabs. These would be promising candidates for achieving highly flexible and robust platforms for integrated optical micro-circuitry.

\begin{acknowledgments}
This work was supported by the EPSRC DTG Grant KD5050. M.F. also acknowledges support from the University of Surrey FRSF, Santander and IAA awards.
\end{acknowledgments}

\end{document}